\documentclass[a4paper,preprint,floatfix]{revtex4}
\usepackage{hyperref}
\usepackage[english]{babel}
\usepackage[latin1]{inputenc}
\usepackage{amsmath}
\usepackage{amssymb}
\usepackage{bm}
\usepackage{graphicx}
\usepackage{setspace}
\usepackage[sort&compress]{natbib}

\newcommand{\Sl}{^\circ\mathrm{S}}
\newcommand{\Nl}{^\circ\mathrm{N}}
\newcommand{\Wl}{^\circ\mathrm{W}}
\newcommand{\El}{^\circ\mathrm{E}}
\newcommand{\dd}{\mathrm{d}}
\newcommand{\Sv}{\,\mathrm{Sv}}
\newcommand{\m}{\,\mathrm{m}}

\begin{document}

  \title{
  Sensitivity of the Atlantic Meridional Overturning Circulation to South Atlantic freshwater anomalies.
  }
  \author{Andrea A. Cimatoribus}
  \email{cimatori@knmi.nl}
  \affiliation{Royal Netherlands Meteorological Institute, De Bilt, The Netherlands}
  \author{Matthijs den Toom}
  \affiliation{Institute for Marine and Atmospheric research Utrecht, Utrecht University, Utrecht, The Netherlands}
  \author{Sybren S. Drijfhout}
  \affiliation{Royal Netherlands Meteorological Institute, De Bilt, The Netherlands}
  \author{Henk A. Dijkstra}
  \affiliation{Institute for Marine and Atmospheric research Utrecht, Utrecht University, Utrecht, The Netherlands}

  \begin{abstract}
    The sensitivity of the Atlantic Meridional Overturning Circulation (AMOC) to changes in basin integrated net evaporation is highly dependent on the zonal salinity contrast at the southern border of the Atlantic.
    Biases in the freshwater budget strongly affect the stability of the AMOC in numerical models.
    The impact of these biases is investigated, by adding local anomaly patterns in the South Atlantic to the freshwater fluxes at the surface.
    These anomalies impact the freshwater and salt transport by the different components of the ocean circulation, in particular the basin--scale salt--advection feedback, completely changing the response of the AMOC to arbitrary perturbations. 
    It is found that an appropriate dipole anomaly pattern at the southern border of the Atlantic Ocean can collapse the AMOC entirely even without a further hosing.
    The results suggest a new view on the stability of the AMOC, controlled by processes in the South Atlantic.
  \end{abstract}

  \maketitle

  \section{Introduction}
  \label{sec:Introduction}

  The instability of the AMOC is often invoked to explain the paleoclimatic evidence of abrupt climate change in North Atlantic areas~\cite{ganopolski_rapid_2001,timmermann_coherent_2003}.
  Also, the possibility of a collapse of the AMOC, triggered by melt water discharge due to global warming, has been suggested~\cite{lenton_tipping_2008}. 
  These hypotheses originate from the seminal work of~\textcite{stommel_thermohaline_1961}, who suggested a view of the AMOC as a two--dimensional circulation between a tropical and a polar box, driven by density differences between the boxes.
  In this simple view, multiple equilibria under the same boundary conditions can exist: an ``ON'' AMOC, dominated by the dense water production in the north of the basin (due to the strong heat flux) and a reverse ``OFF'' circulation with downwelling in the tropical areas (due to the strong evaporation).
  The Stommel model captures the central non--linearity involved: the present--day AMOC  is maintained by  a positive ``salt--advection feedback'' between the northward transport of high salinity waters in the upper ocean and the salinity (and, as a consequence, density) of northern North Atlantic waters.

  Although an obvious oversimplification of the real system, Stommel's results have provided a minimal explanation for the hysteresis behaviour of the AMOC in several other models of higher complexity.
  An abrupt collapse of the AMOC in response to a quasi--equilibrium increase in freshwater forcing in the North Atlantic, not followed by any spontaneous recovery, has been reported in various ocean and climate models of different complexity~\cite{rahmstorf_thermohaline_2005,hawkins_bistability_2011}. 
  These results are not conclusive, though, as they either come from ocean--only and simplified atmosphere models \cite{rahmstorf_thermohaline_2005} or from a coarse--resolution coupled GCM \cite{hawkins_bistability_2011}, leaving open the questions on the role of ocean--atmosphere feedbacks and of the presence of ocean eddies.
  Concerning the ocean--atmosphere interaction, in \textcite{schmittner_model_2005} no abrupt collapse of the AMOC is observed in response to increasing $\mathrm{CO_2}$ concentrations.
  These findings are not directly comparable with those using a direct (and stronger) freshwater forcing in the North Atlantic, but still raise the question whether the multiple equilibria regime of the AMOC is an artifact of simpler climate models, or of ocean--only models.
  A different perspective on this debate is given in~\textcite{huisman_indicator_2010} using an ocean--only model and \textcite{drijfhout_stability_2011} analysing the CMIP3 archive coupled climate models.
  In both works, the existence of a permanent shut--down state of the AMOC is connected with the salt/freshwater budget of the Atlantic Ocean.
  In particular, \textcite{drijfhout_stability_2011} suggest that the lack of a collapsed state in IPCC--class climate models may be attributed to a bias in the salt transport by the AMOC, in connection with the salt--advection feedback at the Atlantic basin scale.

  The value of the net equivalent freshwater import, or scaled salt export by the overturning circulation at the southern border of the Atlantic Ocean (usually taken to be $30\Sl$), shorthanded $M_{ov}$, is the key quantity that signals the coexistence of two stable equilibria of the AMOC.
  A short summary is provided below, with details in the Appendix~\ref{sec:Appendix}; for a complete discussion of the subject of the equivalent freshwater budget of Atlantic Ocean we refer to~\textcite{drijfhout_stability_2011}.
  The following definitions will be used: for a generic field $f$ the barotropic operator is $\overline{f} = \int f \dd z / \int \dd z$, the baroclinic operator is $\tilde{f}=f-\overline{f}$, the zonal operator is $\left<f\right>=\int\limits_{60\Wl}\limits^{20\El} f \dd x/\int \dd x$ and the azonal operator is $f'=f-\left< f \right >$.
  Using the above notation, the equivalent freshwater import (i.e.~the salt export) by the overturning circulation is defined as:
  \begin{equation}
    M_{ov} = -\frac{1}{S_0}\int\limits_{30\Sl} \tilde{v}  \,\left(\left<S\right>-S_0\right) \dd x \dd z,
    \label{eq:Mov}
  \end{equation}
  where $v$ is the meridional velocity ($\tilde{v}$ is thus the baroclinic meridional velocity), $S$ is the salinity and $S_0= \overline{\left<S\right>}_{30\Sl}$.
  If $M_{ov}$ is positive, the overturning circulation is exporting salt out of the Atlantic basin and only the present day ``ON'' state of the overturning is stable. 
  In other terms, a {\em positive} $M_{ov}$ corresponds to a {\em negative} salt--advection feedback at the basin scale, i.e.~more salt is accumulated in the Atlantic Ocean if the AMOC transport decreases.
  If $M_{ov}$ is negative, on the other hand, salt is imported into the basin by the overturning, and a second stable state of the AMOC exists, with reversed or no overturning in the Atlantic Ocean.

  The equivalent freshwater import by the gyre is measured by the azonal part of the equivalent freshwater import, $M_{az}$:
  \begin{equation}
    M_{az} = -\frac{1}{S_0}\int\limits_{30\Sl} v' S' \dd x \dd z.
    \label{eq:Maz}
  \end{equation}
  The complete equivalent freshwater budget for the Atlantic basin is given  by 
  \begin{equation}
   EPR = M_{ov} + M_{az} + M_d + M_{BS} - Q_t - V_t + Res,
   \label{eq:EPR}
  \end{equation}
  where $EPR$ is the basin net evaporation.
  $M_d$ is the transport of equivalent freshwater by subgrid--scale processes (e.g., eddies) at the southern border and $M_{BS}$ is the sum of total volume transport and total equivalent freshwater transport through Bering Strait.  The quantities $Q_t$ and $V_t$ are the basin integrated equivalent freshwater content drift rate due to changes in total salt content and due to changes in volume and sea--ice, respectively. The last term, $Res$, is a residual term that closes the budget.
  It arises due to technical limitations in the determination of $Q_t$ and $V_t$, unavoidable when computing these quantities from model output. The last five terms of Eq.~\ref{eq:EPR} are generally much smaller than the others.
  Eq.~\ref{eq:EPR} states that changes in the net evaporation of the Atlantic Ocean must be mainly compensated by the equivalent freshwater (salt) transport by the baroclinic circulation at the southern border.
  In particular, if $EPR$  remains constant, variations in the equivalent freshwater (salt) transport by the gyre will tend to be compensated by the equivalent freshwater (salt) transport of the overturning and vice versa.

  While the role played by $M_{ov}$ in determining the existence of multiple equilibria of the overturning circulation has been extensively studied in~\textcite{huisman_indicator_2010} and \textcite{de_vries_atlantic_2005}, the importance of the zonal salinity contrast in the South Atlantic, determining $M_{az}$, has received little attention.
  Its relevance has been suggested in \textcite{de_vries_atlantic_2005} where the compensation between $M_{ov}$ and $M_{az}$ has been exploited to control $M_{ov}$ through a small change in the zonal salinity contrast in the subtropical South Atlantic.
  The recent work of~\textcite{drijfhout_stability_2011} has pointed again to the possible impact of the east--west salinity contrast in the South Atlantic for the stability of the AMOC.
  The present study is aimed at further understanding this particular issue: its main aim is to explore the relationship between $M_{ov}$ and $M_{az}$ in a systematic way, and to determine whether changes in the zonal salinity contrast in the South Atlantic can affect AMOC stability.
  The rationale of the experiments and the models used are described in section~\ref{sec:Methods} and the main results in section~\ref{sec:Results}.
A discussion and conclusions follow in section~\ref{sec:discussion}. 

  \section{Model experiments: methods}
  \label{sec:Methods}

\begin{figure*}[htbp]
    \begin{center}
      \includegraphics[width=1.00\linewidth]{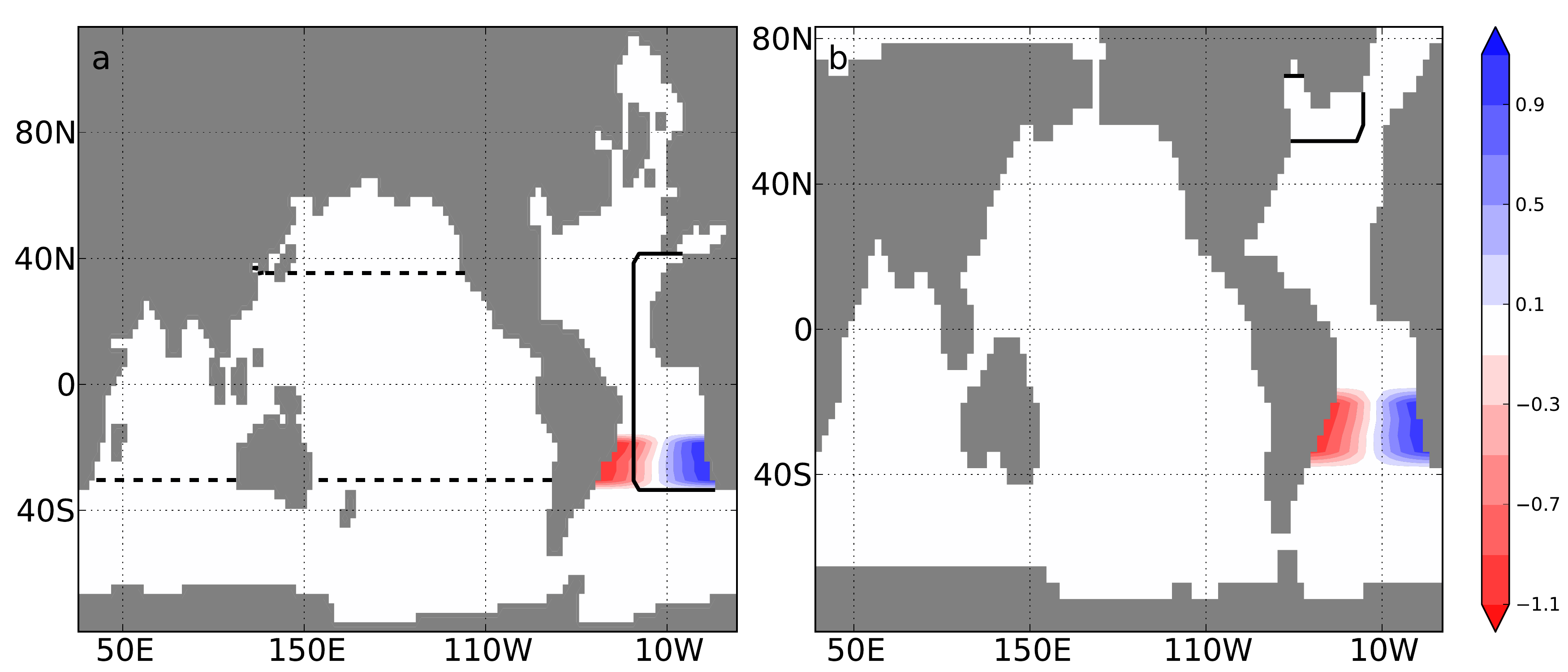}
    \end{center}
    \caption{Areas where the freshwater flux anomalies DIPO and EVAP are implemented in SPEEDO (a) and THCM (b).
    The original model grids have been used (distorted in the North Atlantic and Mediterranean Arctic in the case of SPEEDO).
    The colour shading is the normalised intensity of the DIPO anomaly, the solid contour marks the area where EVAP is implemented, the dashed contour shows for SPEEDO the area used for the compensation of the Atlantic EPR reduction over the Pacific Ocean. For THCM, EVAP anomaly is compensated homogeneously over the rest of the ocean. Positive values correspond to an increased freshwater flux into the ocean.
    }
    \label{fig:regions}
  \end{figure*}

  In this study, we use three different models: the Hybrid Coupled Model (HCM) SPEEDO, the Earth Model of Intermediate Complexity (EMIC) SPEEDO and the fully implicit model THCM (both described in subsections below).
  In the experiments with these models, the effect on the AMOC strength of two different freshwater anomaly patterns is investigated.
  One is a dipole freshwater flux (DIPO) pattern (with amplitude $\delta_p$) applied over the southern part of the South Atlantic (Fig.~\ref{fig:regions}).
  This freshwater anomaly is aimed at studying the effect of different zonal salinity contrasts in  this area on the AMOC.
  The anomaly is expected to have  a direct impact on $M_{az}$, but will also affect $M_{ov}$, in particular when $EPR$ remains  approximately  constant, which is the case as long as the AMOC does not drastically change (collapses). 
  
  The second freshwater anomaly (EVAP)  pattern (with amplitude  $\gamma_p$) changes the net evaporation of the Atlantic basin and is compensated by an opposite anomaly over the tropical Pacific and Indian oceans (Fig.~\ref{fig:regions}).
  This pattern changes the net evaporation and as a result is expected to change $M_{ov}$, with  $M_{az}$ remaining relatively unaffected.
  This anomaly pattern turns out to be an effective control parameter for the AMOC strength, as will be shown below.
  The precise region where this anomaly is applied does not change the sign of the sensitivity of the AMOC to $\gamma_p$, as demonstrated in~\cite{den_toom_effect_2011}, as long as it is applied inside the Atlantic basin.
  When the change in net evaporation is applied closer to the sinking regions of North Atlantic, it has a larger impact in magnitude (but with the same sign) than when the change is applied further south.
  Hence two different areas for applying $\gamma_p$ were chosen in the two models used for this study. 

  \subsection{SPEEDO}

  \subsubsection{SPEEDO EMIC}

  In this work the EMIC SPEEDO~\cite{severijns_efficient_2009} has been used to construct the HCM (see next subsection) and to validate the HCM results.
  The EMIC SPEEDO is an intermediate complexity coupled atmosphere/land/ocean/sea--ice general circulation model with fully resolved ocean and atmosphere dynamics, but simplified physics in its atmospheric component. 
  The ocean component (CLIO) has a horizontal resolution of approximately $3^\circ$ and 20 unevenly spaced vertical levels.
  Convective adjustment is used to avoid static instability in the water column.
  The LIM sea--ice model \cite{graham_sea_1987} is included.
   The atmospheric component of the EMIC is an atmospheric GCM, having a horizontal spectral resolution of T30 and 8 vertical density levels.
   Simple (linearised) parametrisation are included for large--scale condensation, convection, radiation, clouds and vertical diffusion.

  \subsubsection{SPEEDO HCM}

  The HCM SPEEDO~\cite{cimatoribus_global_2011,goosse_importance_1999} includes the same ocean and sea--ice model as the EMIC, forced at the surface by a statistical atmospheric model that, in this case, consists of linear regressions of atmospheric fluxes to the Sea Surface Temperature ($SST$).
  The HCM has been constructed from data of the EMIC SPEEDO. 
  This model is used for most of the experiments.
  A complete description of the definition and test of the HCM can be found in~\textcite{cimatoribus_global_2011}.

  It is forced by a daily climatology for heat, freshwater and momentum fluxes and additionally includes a basic representation of the ocean--atmosphere interaction.
  At each time step linear perturbation terms are calculated, derived from the fully coupled model data and depending on $SST$.
  They are introduced to mimic the effect of atmospheric feedbacks on both local and large scale.
  The local perturbations are a minimal representation of ocean--atmosphere interactions in a statistical steady state.
  The large scale term represents the response of the surface fluxes to changes in meridional overturning circulation strength.
  These perturbation terms reproduce the changes in the surface fluxes connected with e.g.~changes in convection, wind intensity and direction, runoff etc.~\cite{cimatoribus_global_2011}.
  This model design benefits from the fact that the atmosphere, on sufficiently long time scales, can effectively be treated as a ``fast'' component that adjusts to the ocean anomalies.
  The use of a minimal atmosphere renders integrations of tens of thousand years feasible with modest computational requirements.

  The equilibrium solution of SPEEDO, both for the EMIC and the HCM, consists of an Atlantic basin integrated net evaporation overestimated both with respect to most other models and to the few available observations ($0.6\Sv$ in the model as compared to the recent estimate of $0.28\pm0.04\Sv$ by \textcite{talley_freshwater_2008}).
  Furthermore, the zonal gradient of salinity in the South Atlantic is reversed, with a maximum on the eastern side.
  These problems are inherited in the HCM, that features very similar biases.

  \subsubsection{Experimental details}

  The high evaporation over the basin, combined with the low salt export by the gyre due to the reversed zonal salinity profile, force the overturning circulation to export salt ($M_{ov}=0.29 \Sv$) to close the budget.
  As proposed in~\cite{de_vries_atlantic_2005} and~\cite{huisman_indicator_2010}, this situation is connected with the presence of a single equilibrium of the thermohaline circulation, as salt export by the overturning circulation is associated with a negative salt--advection feedback at the basin scale.
  In \textcite{de_vries_atlantic_2005} and~\textcite{cimatoribus_global_2011}, small freshwater corrections were successfully used to change the sign of the salt--advection feedback.
  The anomaly patterns in the two references above are defined in a similar way, even though the sign is opposite there; here anomalies have the same sign as in~\textcite{cimatoribus_global_2011}.
  
  The regions for the flux anomaly patterns are i) the South Atlantic between $30\Sl$ and $20\Sl$ for the DIPO  pattern, which is centred at the zonal midpoint of the basin, and is positive to the east (that is, increased freshwater flux into the ocean in the east) and (ii) the part of the basin east of $20\Wl$, south of Gibraltar Strait and north of the southern tip of Africa for the EVAP pattern (see Fig.~\ref{fig:regions}a).
  The choice of area (ii) is based on the fact that this is the part of the basin where salinity is overestimated the most.
  It may be argued that this choice of EVAP  has a dipole component too, and may thus project on DIPO.
  It will be shown that this does not affect the results substantially; they are comparable with results obtained using an EVAP pattern with no dipole component at all.
  The freshwater anomalies are always implemented as a virtual salt flux at the surface, in order not to influence the heat budget of the basin.
  In SPEEDO the anomalies will always have the same sign, implying that the control run of SPEEDO is taken as the extreme case of highest net evaporation and strongest zonal salinity contrast bias.
  Through the freshwater anomalies these model biases are compensated and even reversed.
  This guarantees that an area of the parameter space is explored that includes the present day state of the ocean.

  Table~\ref{tab:exps} summarises the experiments performed with this model, listing the starting and ending values of the integrated freshwater flux due to each of the two anomaly patterns.
  The experiments are conducted with the following procedure.
  First, in the spin--up phase, the model is brought to a statistical steady state, keeping the freshwater anomalies constant at the initial value listed in Table~\ref{tab:exps}.
  For instance, in experiments A, B or HCMc both anomalies in the initial state are set to zero.
  For C or D, instead, one of the two is kept at a nonzero constant value in the spin--up phase.
  The spin--up time, needed to obtain an ocean in statistical equilibrium, is in the order of thousand years.
  For the reversed experiments Crev and Drev, the initial state is the end state of C and D respectively.
  During the main run, the flux anomalies are changed linearly in time, towards the final value.
  All runs last 12,000 years, except for HCMc and GCMc, lasting 4,000 years.
  These last two runs are meant for checking the consistency of the results from the SPEEDO HCM and the fully coupled model.
  The shorter length of the integration is due to the much higher computational requirements for the fully coupled EMIC.

  The rate of change of the freshwater anomalies, approximately $0.04 \Sv$ per thousand years for the long runs, is far from the values needed to correctly approximate the bifurcation diagram of the AMOC (less than $0.001 \Sv$ per thousand years), in particular close to a bifurcation point~\cite{rahmstorf_bifurcations_1995,timmermann_mechanisms_2005}.
 The focus is here not on the determination of the exact position of the bifurcation points, but rather on the qualitative aspects of the presence/absence of an abrupt collapse of the AMOC for different patterns and amplitudes of the freshwater anomalies applied.
 Despite this deficiency of the experimental setup, it will be shown that the results are consistent with a model (THCM) explicitly solving the steady state problem  for the large scale circulation in the ocean. 

  \subsection{THCM}
  \subsubsection{Model details}

  The ThermoHaline Circulation Model (THCM, \cite{Weijer_etal_03}) is a fully--implicit ocean--only model that is designed to perform numerical bifurcation analysis of the large--scale circulation.
  The model is based on the rigid--lid primitive equations and includes a realistic global model geometry.
  The horizontal resolution is about $4^{\circ}$, and there are 12 levels, ranging in thickness from $50\,\mathrm{m}$ for the top layer to $950\,\mathrm{m}$ for the bottom layer.
  The configuration used is the same as in~\cite{huisman_indicator_2010}, to which we refer for further details.
  The model has several deficiencies \cite{Dijkstra_Weijer_05} that make it less well suited for quantitative analyses.
  Rather, it allows for an efficient way of exploring qualitatively the properties of the large--scale global circulation.

  In THCM, heat fluxes are determined by a simple two--dimensional energy balance model that is coupled to the upper ocean layer~\cite{Weijer_etal_03}.
  Sea--ice is not included in the model.
  For the wind stress, the annual mean field provided by~\cite{Trenberth_etal_89} is used.
  The reference (unperturbed) freshwater flux is diagnosed from the sea surface salinity restoring to the Levitus~\cite{levitus_world_1994} climatology.
  The zonal salinity contrast in the South Atlantic is hence approximately correct, with basin--integrated net evaporation amounting to $0.3\Sv$.
  In the reference case ($\delta_p = \gamma_p = 0$), the surface loss of freshwater is compensated by freshwater transport by both the AMOC ($M_{ov}=0.10\Sv$) and the azonal transport ($M_{az}=0.05\Sv$); the remainder due to diffusion and transport across the northern border of the Atlantic.
  Azonal transport is weak compared to SPEEDO due to the weak barotropic circulation, so that diffusive transport plays a primary role in closing the equivalent freshwater budget.

  \subsubsection{Experiment details}

  With the aim of building a framework for understanding the SPEEDO HCM results, the sensitivity of THCM to changes in $\delta_p$ and $\gamma_p$ is explored.
  As in \cite{Weijer_etal_03,Dijkstra_Weijer_05,huisman_indicator_2010}, but different from what is done in the HCM experiments, the net evaporation change is achieved by applying an anomalous flux of strength $\gamma_p$ south of Greenland (see Fig.~\ref{fig:regions}b), compensating this flux everywhere else.
  The pattern of the dipole anomaly is similar to the one used in the HCM SPEEDO experiments (Fig.~\ref{fig:regions}a).
  $\gamma_p$ is used as the primary bifurcation parameter and parameter space is explored by constructing bifurcation diagrams for a number of discrete values of $\delta_p$.
  From the bifurcation diagrams, a regime diagram can be constructed, delineating the regions of existence of the present--day and collapsed states of the overturning circulation.

  \section{Model experiments: results}
  \label{sec:Results}

  First, the results from the SPEEDO HCM are discussed, analysing in particular the relationship between the freshwater anomalies and the AMOC strength, and the equivalent freshwater transport by the overturning and azonal components of the circulation.
  Results from the SPEEDO HCM and the EMIC are reported in figures~\ref{fig:AMOC1} to~\ref{fig:30N2}.
  Thereafter, the results will be interpreted in the framework of dynamical systems, using the regime diagram obtained from THCM (Fig.~\ref{fig:THCM}).

  \subsection{SPEEDO HCM}
  \label{sec:Results.SPEEDO}

  The freshwater anomalies  induce a strong overshooting of the AMOC (Figs.~\ref{fig:AMOC1} and~\ref{fig:AMOC2}, black lines).
  Changes in the boundary conditions can trigger convection at new sites, that will then give a substantial contribution to the overturning strength in our low resolution configuration.
  In these experiments, the freshwater anomalies are applied far from the convection sites, so the initial response in the North Atlantic can actually be that of a slight increase in surface density, that in turn triggers convective adjustment at new grid cells with a consequent AMOC overshooting.
  Even if the magnitude of the increase in overturning strength is overestimated, there is no reason to believe that the sign of the response is wrong.
  Moreover, the overshootings have an intermittent character and the overall trend in AMOC strength is not affected by them.

  \subsubsection{Decreasing EVAP, no DIPO anomaly}

\begin{figure*}[htbp]
    \begin{center}
      \includegraphics[width=1.0\linewidth]{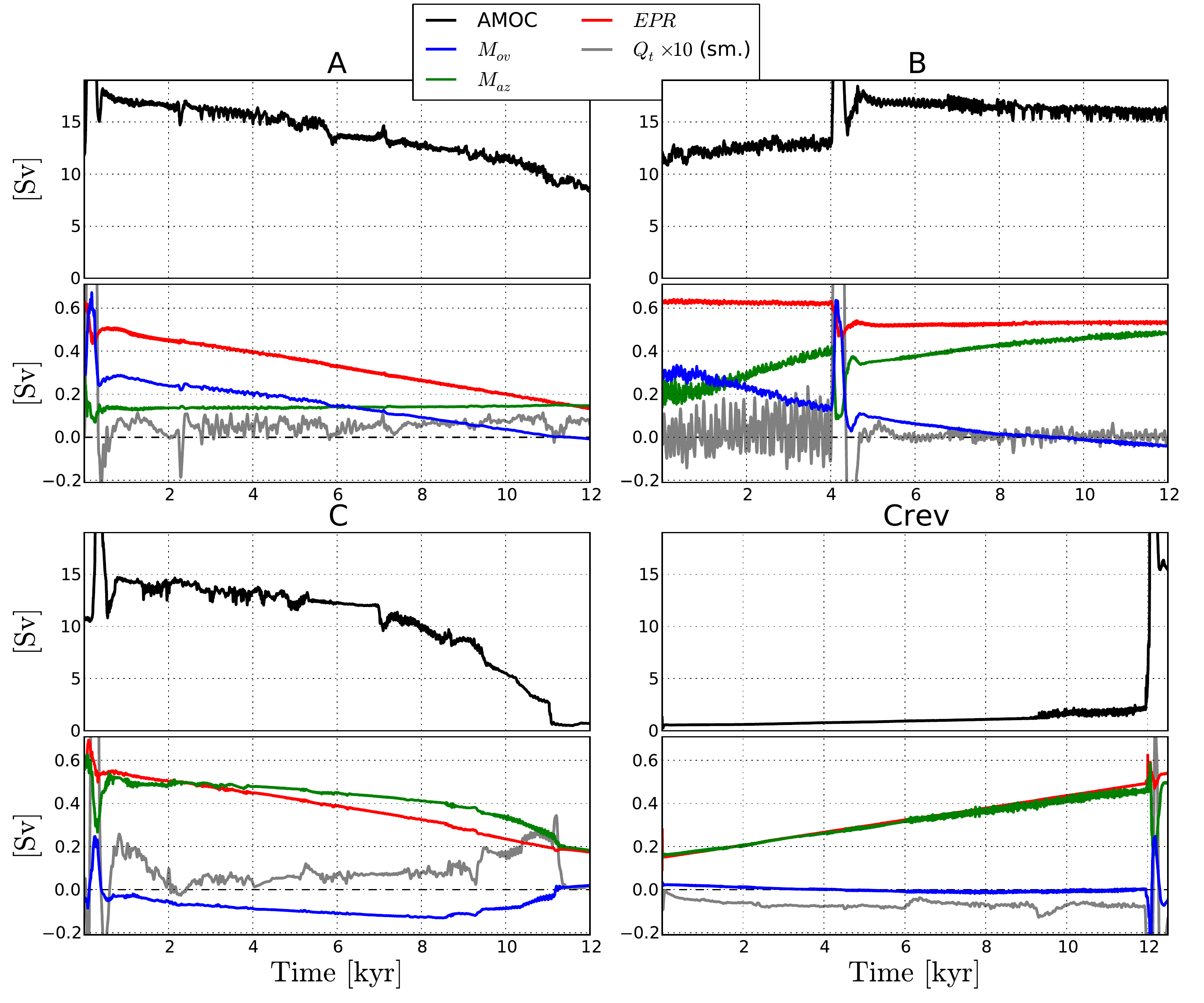}
    \end{center}
    \caption{AMOC and equivalent freshwater budget at the southern border of Atlantic Ocean for experiments A, B, C, Crev.
    For each experiment, two panels are shown: in the upper one the evolution of the maximum of the AMOC streamfunction below $550m$ is shown, in the lower panel the evolution of some of the terms of the equivalent freshwater budget (equation~\ref{eq:EPR}) are reported. $Q_t$ (grey line) is magnified 10 times and smoothed for plotting purposes, through a convolution with a Hanning window 71 years wide.
    This is done to remove fast fluctuations that would otherwise render the plot unreadable.}
    \label{fig:AMOC1}
  \end{figure*}

\begin{figure*}[htbp]
    \begin{center}
      \includegraphics[width=1.0\linewidth]{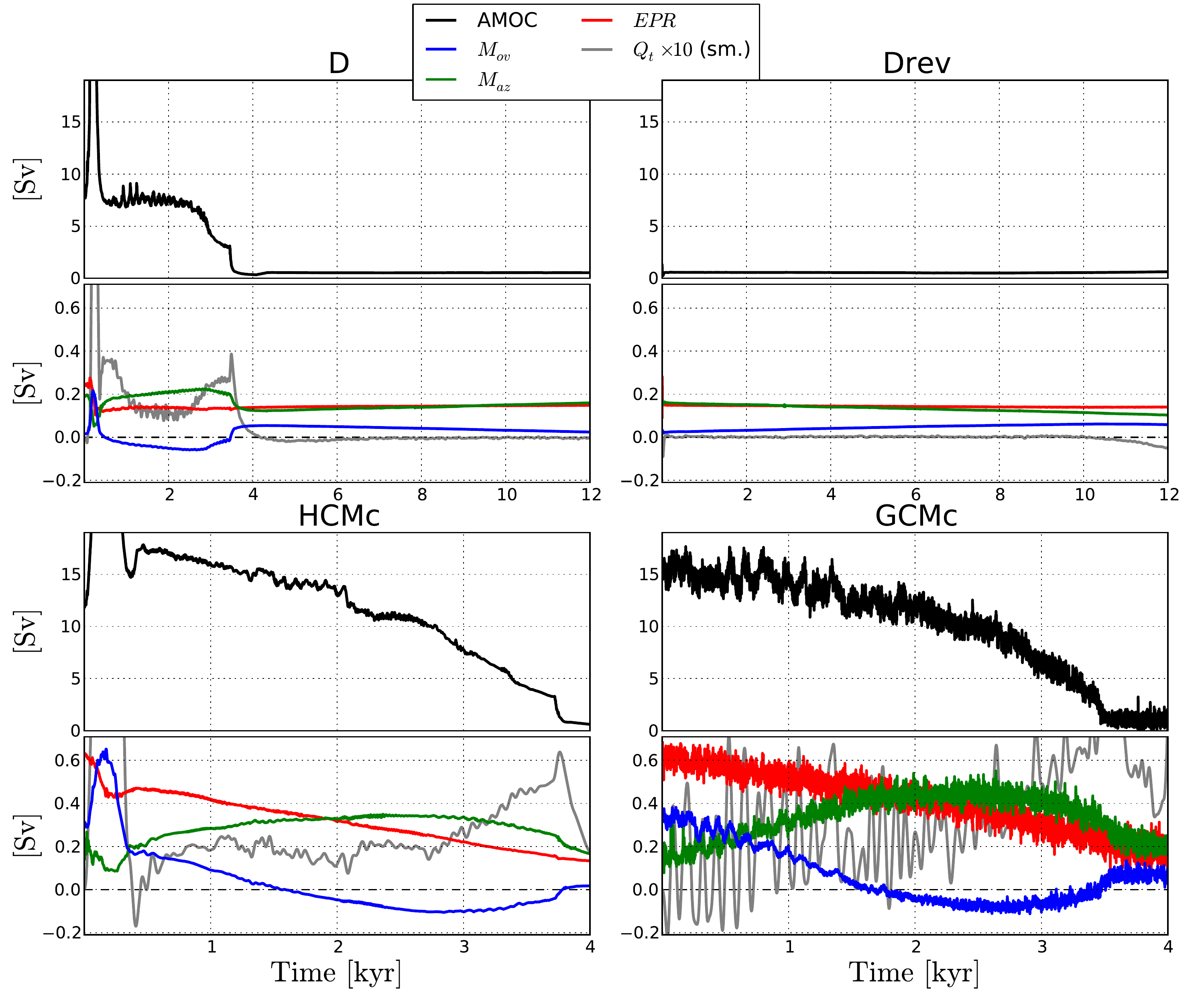}
    \end{center}
    \caption{Same as Fig.~\ref{fig:AMOC1} for experiments D, Drev, HCMc and GCMc.
    \label{fig:AMOC2}
    }
  \end{figure*}

  The first experiment, A, includes no dipole anomaly, but solely a decrease of net evaporation over the Atlantic basin.
  The ocean state of this experiment can be considered as an extreme case of an ocean model with a zonal salinity bias.
  The final value of the evaporation anomaly is $0.4\Sv$, giving a substantial decrease in $EPR$ (Fig.~\ref{fig:AMOC1}A, red line).
  The response of the AMOC, after an initial overshooting, is that of a strong decrease in strength (maximum is halved at the end of the run).
  No abrupt collapse is observed, and the decrease slightly deviates from linear behaviour only after year 10,000 of the simulation.
  $M_{ov}$  (Fig.~\ref{fig:AMOC1}A, blue line) is positive in the initial phase, and approaches zero only at the end of the experiment.
  The change in $EPR$ impacts only $M_{ov}$, leaving the other terms in the equivalent freshwater budget unchanged (the terms of Eq.~\ref{eq:EPR} not plotted in Fig.~\ref{fig:AMOC1} are, to a very good approximation, constant in all experiments (see Appendix~\ref{sec:Appendix})).
  No compensation is seen between $M_{ov}$ and $M_{az}$, and the change in $EPR$ is thus almost completely balanced by a change in $M_{ov}$.
  The change in $M_{ov}$ can be explained by the decrease in overturning strength ($M_{ov}$ scales linearly with the overturning strength) and by the changes in the zonally averaged salinity profile at $30\Sl$ (see Fig.~\ref{fig:S30S}A).
  The latter figure shows that the $EPR$ anomaly affects the deeper branch of the AMOC (freshening) as well as the upper one (becoming saltier).
  The net effect is an increased salt transport into the Atlantic basin by the overturning circulation, consistently with $M_{ov}$ in Fig.~\ref{fig:AMOC1}A.
  The grey line of Fig.~\ref{fig:AMOC1}A shows the rate of change of the equivalent freshwater content in the Atlantic basin ($Q_t$ in Eq.~\ref{eq:EPR}, multiplied by a factor 10).
  $M_{ov}$ does not compensate exactly for the change in $EPR$, as $Q_t$ is greater than zero. 
  
  The basin is thus freshening, suggesting that this is the main cause for the decrease in AMOC strength.
  As the strength of the AMOC is determined by the amount of sinking in the high latitudes of the North Atlantic, the equivalent freshwater transport in the basin is also diagnosed at $30\Nl$ (Fig.~\ref{fig:30N1}) \footnote{As the model grid is distorted in North Atlantic and Arctic Oceans, the latitude of this transect is only approximate. The model grid is needed in order to avoid interpolation errors.}.
  The AMOC strength seems to be controlled by the overturning component of the salt transport at this latitude (shorthanded $M_{ov}^{30N}$, as it is defined in the same way as $M_{ov}$ at the southern border of the basin).
  This quantity is negative, since at this latitude the salt transport is always northward.
  The $EPR$ reduction makes $M_{ov}^{30N}$ less negative.
  Even if this change is almost fully compensated by the azonal part of the transport at $30\Nl$, the decreased salt transport to higher latitudes brings an effective freshwater perturbation in the sinking regions, as shown by the positive value of $Q_t^{30\Nl}$ (multiplied by a factor 5, grey line in Fig.~\ref{fig:30N1}A), i.e.~the drift in equivalent freshwater content north of $30\Nl$ in the Atlantic and Arctic.
  $EPR$ north of $30\Nl$ ($EPR^{30\Nl}$, red line in Fig.~\ref{fig:30N1}A) remains on the other hand almost constant.
  The deviation from the linear decrease of the AMOC appears after the azonal transport at $30\Nl$ ($M_{az}^{30N}$) becomes negative.
  This evidence, confirmed by the other experiments, suggests that a qualitative change in the AMOC response takes place when the overturning circulation is not able to import enough salt from the tropics into the subpolar region to compensate for precipitation change over the area.

  \subsubsection{Increasing DIPO, no EVAP anomaly}

  In experiment B, $EPR$ is unchanged, while the dipole anomaly increases from $0$ to $0.5\Sv$ during the run.
  In response to this, the AMOC slightly increases until year 4,000.
  After a sharp rise between years 4,000 and 4,600, the AMOC starts to decrease until the end of the simulation.
  $M_{ov}$ and $M_{az}$ at the southern border change with opposite sign in this case, and approximately compensate each other (Fig.~\ref{fig:AMOC1}B).
  This is connected with the fact that, in the budget of Eq.~\ref{eq:EPR}, all terms remain approximately constant, with the exception of $M_{az}$, controlled by the dipole anomaly, and $M_{ov}$, opposing the azonal term, while $EPR$ remains almost constant.
  By means of this compensation mechanism, the sign of $M_{ov}$ changes during the experiment.
  The total salinity content of the basin is not significantly affected by the changes in the dipole anomaly ($Q_t \approx 0$ in the time mean, see figure~\ref{fig:AMOC1}B).
  This is the principle exploited in the experiments of~\textcite{de_vries_atlantic_2005} and \textcite{cimatoribus_global_2011}, where a dipole anomaly was applied to change the sign of $M_{ov}$.
  Because of the compensation between $M_{ov}$ and $M_{az}$, the dipole anomaly has to induce changes in zonally averaged salinity as well on the longer time scale.
  Figure~\ref{fig:S30S}B shows that the main effect of the dipole anomaly on the zonally averaged salinity is to increase the salinity in the upper branch of the AMOC (between approximately $250$ and $1000\m$).
  Virtually no changes are observed in the lower branch (approximately between $1000$ and $2500\m$) before and after the overshoot, but during the overshoot a significant freshening takes place, reflecting changes in deep convection.
  The dipole anomaly is changing $M_{ov}$ modifying the intermediate waters, whose salinity increases in response to the increased salinity export by the azonal part of the circulation (salinity increases not only in south--western Atlantic, but also more to the south).
  During the whole run the dipole anomaly shifts the salt transport between different terms of the transport, but the overall salt content of the Atlantic Ocean is not changed, nor is $EPR$ (again, excluding the discrete jump during the overshooting).
  Also in this case, changes in AMOC strength are controlled by the equivalent freshwater transport by the overturning at $30\Nl$ (AMOC increases when $M_{ov}^{30\Nl}$ decreases, and vice versa, see Fig.~\ref{fig:30N1}B). 

  \subsubsection{Changing EVAP, constant DIPO anomaly}

\begin{figure*}[htbp]
    \begin{center}
      \includegraphics[width=1\linewidth]{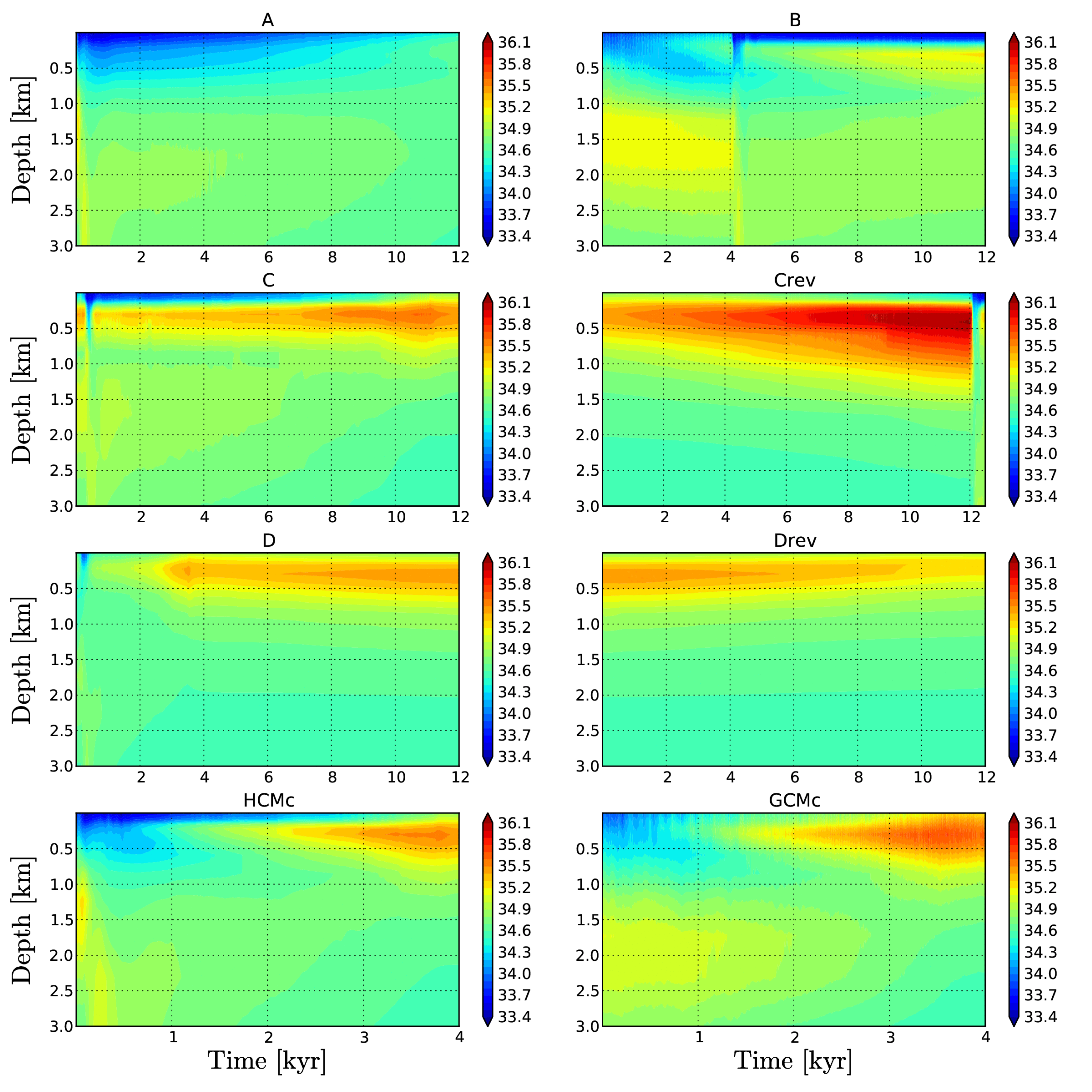}
    \end{center}
    \caption{Temporal evolution of the zonal average of salinity between $60\Wl$ and $20\El$ at the southern border of the Atlantic Ocean as a function of time and depth. 
    \label{fig:S30S}
    }
  \end{figure*}

\begin{figure*}[htbp]
    \begin{center}
      \includegraphics[width=1.0\linewidth]{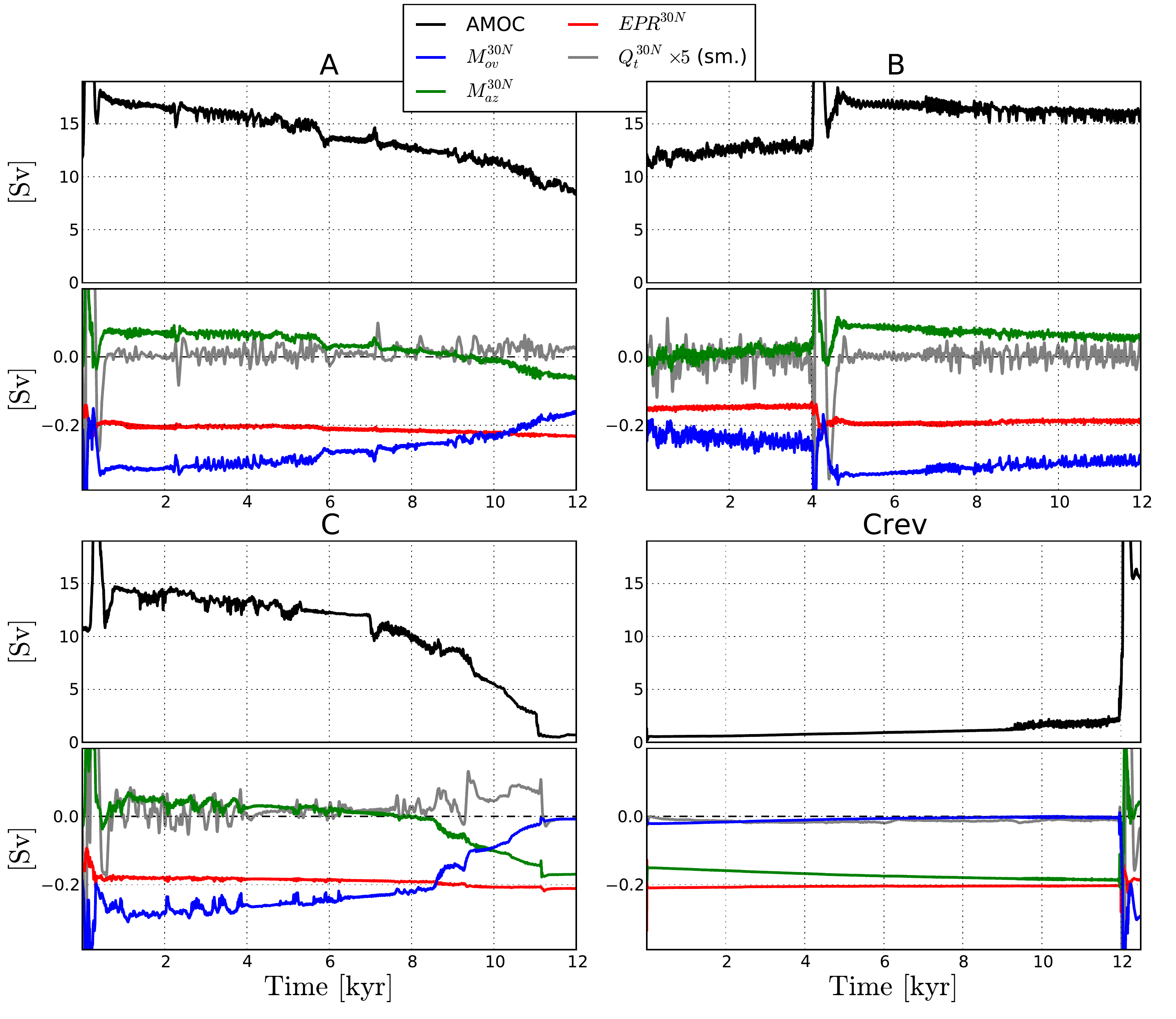}
    \end{center}
    \caption{Same as figure~\ref{fig:AMOC1}, but considering a section at approximately $30\Nl$, and the portion of the Atlantic and Arctic basin northwards.
    \label{fig:30N1}
    }
  \end{figure*}

\begin{figure*}[htbp]
    \begin{center}
      \includegraphics[width=1.0\linewidth]{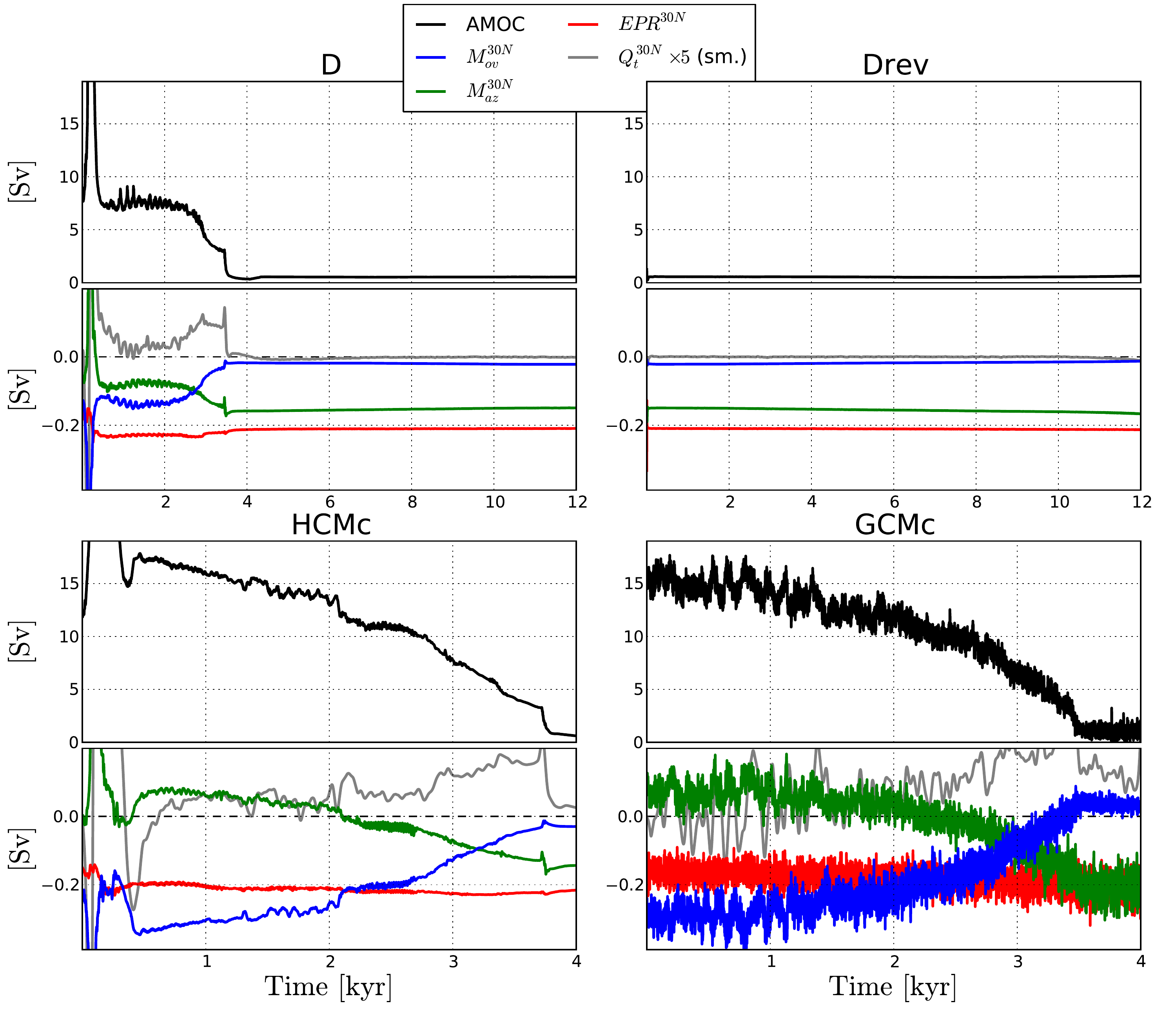}
    \end{center}
    \caption{Same as figure~\ref{fig:30N1} for experiments D, Drev, HCMc and GCMc.
    \label{fig:30N2}
    }
  \end{figure*}

  In experiment C, the same $EPR$ reduction as in A ($0.4\Sv$) was applied to an ocean state that already includes a dipole anomaly of $0.5\Sv$.
  The zonal salinity contrast in the South Atlantic is reversed compared to experiment A (salinity maximum in the west of the basin instead of the east).
  The response of the AMOC is now totally different.
  The initial strength (discarding the initial overshooting) is just $2\Sv$ lower than in experiment A, but in experiment C the $EPR$ reduction causes the AMOC to completely collapse at year 11,000.
  The initial (linear) response of the AMOC in C is not stronger than in A, but the linear behaviour breaks down near approximately year 8,000, with a faster linear decrease at first, and then a final complete collapse.
  Thus, the AMOC may collapse by freshwater anomalies applied far from the sinking regions.
  This is a clear evidence of the fact that the sign of the sensitivity of the AMOC to freshwater anomalies, on longer time scales, is not dependent on the area where the anomaly is applied~\cite{den_toom_effect_2011,hawkins_bistability_2011}.
  Also the behaviour of $M_{ov}$ and $M_{az}$ is totally  different from that in experiment A.
  In the initial state, $M_{ov}$ is slightly negative, and tends to decrease as long as the linear response of the AMOC is maintained.
  Both the zonal and azonal components of the transport are affected by the reduction in $EPR$, and both contribute to closing the equivalent freshwater budget of the basin.
  As in A, the net $EPR$ change affects both the upper and the lower branches of the AMOC, but the zonally averaged salinity profile is very different in C (compare Fig.~\ref{fig:S30S} panels A and~C).
  The constant dipole anomaly applied in C induces a reduction in contrast between the upper and lower branches of the AMOC. 
  As in A, $M_{ov}$ decreases when $EPR$ decreases, but the lower salinity contrast reduces the ability of the AMOC to import more salt into the basin as $EPR$ decreases.
  The budget must be closed by a decrease in $M_{az}$ as well.
  The changes in the equivalent freshwater transport do not fully compensate the reduction in $EPR$, and the basin is freshening more than in experiment A.
  This can be seen by comparing the grey lines in figure~\ref{fig:AMOC1}A and C, showing that $Q_t$ is more positive in the latter experiment.
  In correspondence with the break of the linear response of the AMOC, the relationship between $M_{ov}$ and $M_{az}$ is suddenly modified, and their changes start to be of opposite sign ($M_{az}$ keeps decreasing, but $M_{ov}$ starts to increase).
  This increase ($M_{ov}$ becomes less negative) is due to the quickly weakening AMOC.
  The decreased salt import by the AMOC brings an effective freshwater anomaly in the basin.
  This is the manifestation of the basin scale salt--advection feedback.
  The decreasing intensity of the AMOC amplifies the initial perturbation through $M_{ov}$ which leads to the collapse.
  Considering the section at $30\Nl$ (Fig.~\ref{fig:30N1}C), the situation is very similar to experiment A, but the subpolar Atlantic is freshening much more in this case (grey line).
  As for experiment A, the initial linear response of the AMOC is broken down when $M_{az}^{30N}$ becomes negative.

  The reversed experiment, Crev, starts from the end of experiment C and consists of a reduction of the $EPR$ anomaly from $0.4\Sv$ back to zero (i.e.~an increasing $EPR$) over the Atlantic Ocean.
  The changes in the net evaporation of the basin are compensated only by $M_{az}$ (Fig.~\ref{fig:AMOC1}Crev), as the (reverse) overturning is weak and shallow.
  This compensation is incomplete, and the salinification of the basin (grey line in Fig.~\ref{fig:AMOC1}Crev) leads to a recovery at the end of the run (a few hundred years after year 12,000).
  The area where the two equilibria exist seems to extend over almost the whole parameter range covered by experiments C and Crev.
  It must be kept in mind, however, that the boundaries of the multiple equilibria region are likely to be overestimated in these transient hysteresis experiments.
  The fact that the anomalies change too fast for an equilibrium to be maintained will tend to delay the transitions between the two states, due to the inertia of the system (see e.g.~\cite{kuehn_mathematical_2011}).
  Furthermore, the low noise level in the HCM provides an unrealistically weak source of perturbations to the circulation, that reduces the chances of transitions to occur before the bifurcation points are reached~\cite{meunier_noise_1988}.
  The initially slow recovery is already accompanied by a strong increase of salinity in the upper Atlantic (Fig.~\ref{fig:S30S}Crev), which provides a positive density perturbation eventually triggering the AMOC onset.

  \subsubsection{Changing DIPO, constant EVAP anomaly}

  In the last full length experiment, D, the increase in the dipole freshwater anomaly is applied over an Atlantic Ocean with reduced net evaporation.
  The initial state features an AMOC that, although stable, is markedly weaker than in the previous cases.
  Figure~\ref{fig:30N2}D shows that $M_{az}^{30N}$ is already negative at the beginning of the run, a situation that is connected with an increased sensitivity of the AMOC to $EPR$ perturbations.
  During the experiment, the AMOC shows little sensitivity to the changes in the dipole anomaly before approximately year 3,000.
  After this point, the circulation quickly collapses.
  This experiment shows that the dipole anomaly, that does not affect the integrated net evaporation of the Atlantic basin, is sufficient to collapse the AMOC provided that the basin is not too strongly evaporative.
  Also in this case a compensation between $M_{ov}$ and $M_{az}$ at the southern border is observed, up to the point where the collapse is triggered (Fig.~\ref{fig:AMOC2}D).
  In this case, $M_{ov}$ is negative, and a strong freshening of the basin is observed before the collapse, showing that the overturning circulation is not able to completely compensate the increasing salt export by the azonal component.
  After the collapse is triggered, the basin scale salt--advection feedback is playing a major role in increasing $Q_t$, through the increase in $M_{ov}$ becoming less negative as the AMOC weakens.
  The reversed circulation that establishes after the collapse has a slightly positive $M_{ov}$, that guarantees the closure of the budget, bringing $Q_t$ to zero.

  To check whether this collapse is permanent, the experiment is repeated, decreasing again to $0$ the value of the dipole anomaly (Drev).
  The AMOC shows no sign of recovery, despite the fact that $M_{az}$ and $M_{ov}$ are (slowly) responding to the increasing freshwater anomaly.
  At the end of the run $Q_t$ deviates markedly from zero towards negative values.
  The salinification of the basin is likely a prelude to a recovery for negative values of the dipole anomaly, but the run was not extended into that unrealistic parameter regime.
  
  It is concluded from these experiments that the zonal salinity contrast plays a fundamental role in determining the sensitivity of the AMOC, and a zonal bias in the salinity distribution in the South Atlantic can completely change the sensitivity of the AMOC to perturbations in the basin integrated surface freshwater forcing (experiments A and C).
  In particular, the zonal salinity contrast strongly affects the equivalent freshwater transport by the azonal and overturning circulation, and can determine as a consequence the stability of the AMOC.
  The dipole freshwater anomaly that corrects such a bias does not have a large impact on the AMOC strength, but can cause a collapse in a basin with low $EPR$.

  \subsubsection{EMIC--HCM comparison}

  Last, a comparison between the HCM and the full EMIC is performed, with two runs named HCMc and GCMc respectively.
  Both models start from the initial state of experiment A, and the two freshwater anomalies are then both increased during 4,000 years.
  The response in the two models is very similar (Fig. 2--4), supporting the choice for the HCM in the other runs.
  The main differences are the lower variability in HCMc and the earlier collapse in GCMc.
  The weak variability of the HCM model originates from the model definition itself, which does not include any noise term able to mimic atmospheric variability on short time scales.
  This issue has already been discussed in~\textcite{cimatoribus_global_2011}, and is not affecting the very long time scales considered in this work.
  The collapse for lower values of the anomalies in GCMc can be explained by GCMc having higher noise levels that provides stronger perturbations to the AMOC (higher chances of switching between two equilibria).
  The sensitivity of the overturning strength to the freshwater anomalies, as well as that of the salt transports, is indeed very similar in the two experiments.

  \subsection{THCM}
  \label{sec:Results.THCM}

\begin{figure*}[htbp]
    \begin{center}
      \includegraphics[trim=0cm 0cm 0cm 0.5,clip,width=0.59\linewidth]{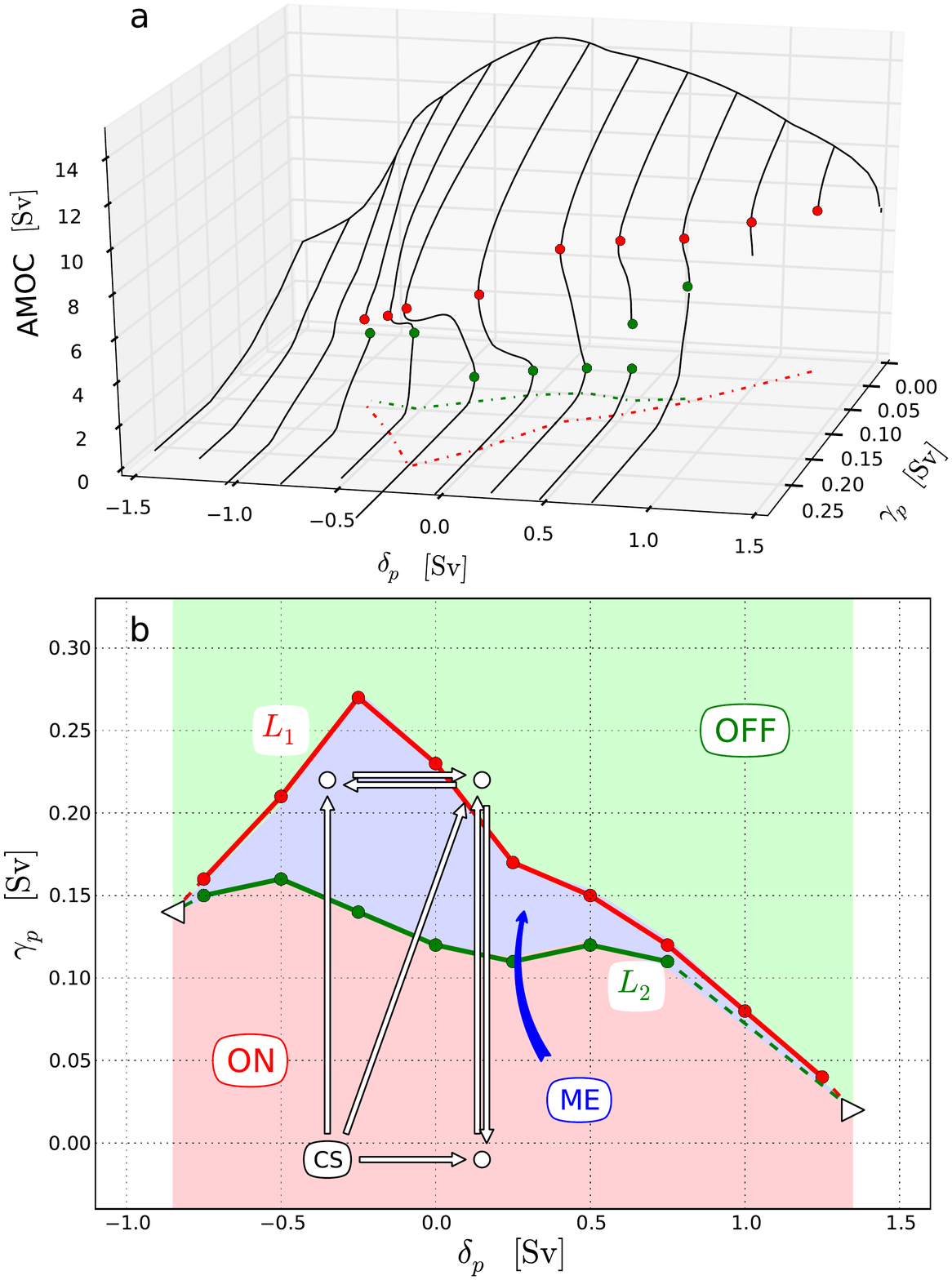}
    \end{center}
    \caption{
    a. Collective three-dimensional plot of bifurcation diagrams from THCM experiments.
    The  maximum AMOC streamfunction is shown as a function of the two control parameters $\delta_p$ (strength of the DIPO anomaly) and $\gamma_p$ (strength of EVAP anomaly).
    Some solution branches are incomplete due to computational problems (see text).
    In both panels, the red dots mark the saddle node at the end of the ``ON'' solution branch ($L_1$); the green dots refer to the saddle node marking the end of the ``OFF'' solution branch ($L_2$).
    The projection of the position of the saddle--node bifurcations on the ($\delta_p$,$\gamma_p$) plane is also shown (enlarged in panel b).
 \\
    b. Regime diagram for the AMOC streamfunction in THCM. 
    The dots mark the positions of the saddle nodes and the lines are guides for the eye.
    The two triangles mark the purely hypothetical positions of the two cusps, where the two saddle nodes merge.
    The red (green) shaded area is the region where only the ``ON'' (``OFF'') solution exists, the blue area marks the multiple equilibria regime (ME).
    Outside the area horizontally bounded by the cusps, the two solutions are not separated by an unsteady state, and no abrupt transition is possible changing $\gamma_p$.
    Superimposed to the regime diagram, in white, are the approximate trajectories of the experiments performed with the SPEEDO HCM.
    The end and initial states of SPEEDO HCM experiments are marked by white dots.
    The control state of the SPEEDO HCM (marked ``CS'') is shifted with respect to that of THCM (see text), and all the anomalies applied to it are shifted as a  consequence.
    $\gamma_p$ has different scales for the two models (see text).
    \label{fig:THCM}
    }
  \end{figure*}

  The results of the previous section will be further interpreted with a bifurcation analysis obtained with THCM.
  Due to the substantial differences between the models, the comparison will be mainly qualitative, enabling to interpret the findings reported above in the framework of bifurcation theory.
  The results from THCM are summarised in Fig.~\ref{fig:THCM}, showing first the maximum of the AMOC streamfunction   as a function of the two control parameters $\gamma_p$ and $\delta_p$, i.e.~the integrated intensity of the freshwater anomalies (Fig.~\ref{fig:THCM}a).
  The paths of the saddle--node bifurcations (the points where the abrupt transition between the two states takes place) are plotted as the dotted curves in Fig.~\ref{fig:THCM}a, shown on the bifurcation diagrams as circles and, in more detail, in the regime diagram in Fig.~\ref{fig:THCM}b.
  When $\delta_p$ is changed, two modifications of the bifurcation diagram occur: (i) the saddle--node bifurcations both shift in the same direction along the $\gamma_p$ axis (rigidly shifting the whole multiple equilibria (ME) regime region) or (ii) the two saddle--node bifurcations shift in opposite directions (affecting the width of the ME regime in the $\gamma_p$ direction).
  The path of the one on the ``ON'' branch ($L_1$) is plotted in red and the one on the ``OFF'' state  ($L_2$) in green.
  From this regime diagram, the shift of the ME regime and the changes in hysteresis can be distinguished.

  The estimate of the initial control state of SPEEDO, marked with CS in Fig.~\ref{fig:THCM}b, is based on the surface salinity biases diagnosed in the model, and on the qualitative comparison of the behaviour of the HCM with respect to THCM.
  The trajectories in the phase space of the different experiments performed with the SPEEDO HCM are drawn as white arrows.
  These trajectories are, by necessity, only rough estimates since $\gamma_p$ scales differently in the two models, being applied in different regions.
  The area where two steady states are possible under the same boundary conditions (ME) is marked in blue in the figure.
  This region is bounded by the two saddle node bifurcation lines, that in turn merge in two cusps at sufficiently large values of the dipole anomaly ($\mathcal{O}(1\Sv)$ for the model used here).
  In the vicinity of the two cusps, the computational time needed for finding steady state solutions diverges, and the results are thus missing there; their actual position can only be guessed.
  The AMOC is substantially weakened outside the region bounded by the cusps, for any value of $EPR$ (Fig.~\ref{fig:THCM}a).

  The diagram is not symmetric with respect to the $\delta_p=0$ line: for positive values of the DIPO anomaly ($\delta_p > 0$, i.e., freshwater added in the east) the two saddle nodes slowly approach each other moving towards the cusp, but quickly shift towards the $\gamma_p=0$ line (in particular the first saddle node, red line in Fig.~\ref{fig:THCM}).
For negative values of the DIPO anomaly ($\delta_p < 0$), the bifurcation points approach each other much faster, with an evident shift of the second saddle node towards positive $\gamma_p$. Outside the multiple equilibria regime, only one steady state is available.
  Two distinct ``ON'' and ``OFF'' states are well defined in the ME interval between the two cusps, as in this area an unstable solution is dividing the two states (see e.g.~\cite{huisman_indicator_2010}).
  Changes in the net freshwater anomaly, here, force the system to jump from one solution to the other when a saddle node is reached.
  The transition in the two opposite directions takes place at different values of $\gamma_p$, i.e.~the response of the circulation shows hysteresis.
  Outside this area, a continuous change between the two states, as a function of the net evaporation change, is observed.
  In this region of the diagram, no hysteresis behaviour is possible.
  It must be noted that in a more realistic framework, where variability in the ocean and the atmosphere would provide a source of stochastic perturbations, the area where hysteresis can be detected would shrink, as stochastic perturbations would render the positions of the two saddle nodes practically indistinguishable if close enough~\cite{meunier_noise_1988}.

  With a larger positive dipole anomaly, a smaller $\gamma_p$ is needed to reach the multiple equilibria regime, and eventually to collapse the AMOC.
  This is confirmed by the results from the SPEEDO HCM, experiments A and C (represented by the vertical white arrows in Fig.~\ref{fig:THCM}b).
  When the dipole anomaly is applied (experiment C), the same net evaporation reduction used in A is sufficient to completely collapse the AMOC, crossing the first saddle node $L_1$. 
  In experiment C, the value of $M_{ov}$ is also much lower than in A and it is negative, suggesting that the AMOC in the HCM, with the dipole anomaly applied, is in the multiple equilibria regime, as confirmed by the reversed experiment Crev.

  The strength of the AMOC for $\gamma_p = 0$ only weakly depends  on $\delta_p$, with a maximum close to $\delta_p=-0.25\Sv$   (Fig.~\ref{fig:THCM}a), but the basin integrated net evaporation is the main control parameter of the AMOC strength.
  A similar, weak dependence of the AMOC strength on $\delta_p$ was found in SPEEDO (experiment B).
  Finally, experiment D (upper horizontal arrow) shows how the $\delta_p$ interval for which an ``ON'' state of the overturning circulation is available quickly shrinks for decreasing $EPR$ values (increasing $\gamma_p$), up to the point where a small dipole anomaly at the southern border of the Atlantic (approximately $0.125\Sv$ in experiment D) is sufficient for causing a complete collapse of the AMOC.

  \section{Discussion and conclusions}
  \label{sec:discussion}
  
  In this work, the central importance of the Atlantic equivalent freshwater budget in determining the stability of the AMOC has been demonstrated.
  In particular, it was shown that a correct representation of the zonal salinity contrast at the southern border of the Atlantic is fundamental in determining the stability properties of the overturning circulation and the existence of the ME regime.
  The strength of a freshwater anomaly necessary to collapse the overturning is highly sensitive to biases in the salinity import by the azonal circulation at the southern boundary, in turn mainly determined by the zonal salinity difference.
  It was also shown that a dipole freshwater anomaly summing up to zero, applied over Southern Atlantic, is sufficient to collapse the overturning circulation if the basin net evaporation is sufficiently low.
  These results demonstrate that the azonal transport, connected to the three dimensional wind--driven gyre, plays a major role in controlling the AMOC stability, through the compensation mechanism between salt transport by the overturning and the gyre, whenever the net evaporation over the Atlantic basin remains approximately constant.
  This interplay of the meridional and horizontal circulation challenges the two dimensional view of the AMOC, and once again suggests that the results from Stommel--type models (e.g.~\cite{stommel_thermohaline_1961,rahmstorf_freshwater_1996,park_1999,scott_interhemispheric_1999,johnson_reconciling_2007,guan_stommels_2008}) should be evaluated with great care.
  In particular, the idea of the AMOC stability determined only by salinity in the sinking regions of North Atlantic seems to be an oversimplification of the real system.
  In summary, the salinity in the subpolar North Atlantic is one of the main controls of the strength of the AMOC, but the existence of a stable collapsed state of the AMOC is controlled by the equivalent freshwater budget of the entire basin.

\begin{figure*}[htbp]
    \begin{center}
      \includegraphics[width=1.0\linewidth]{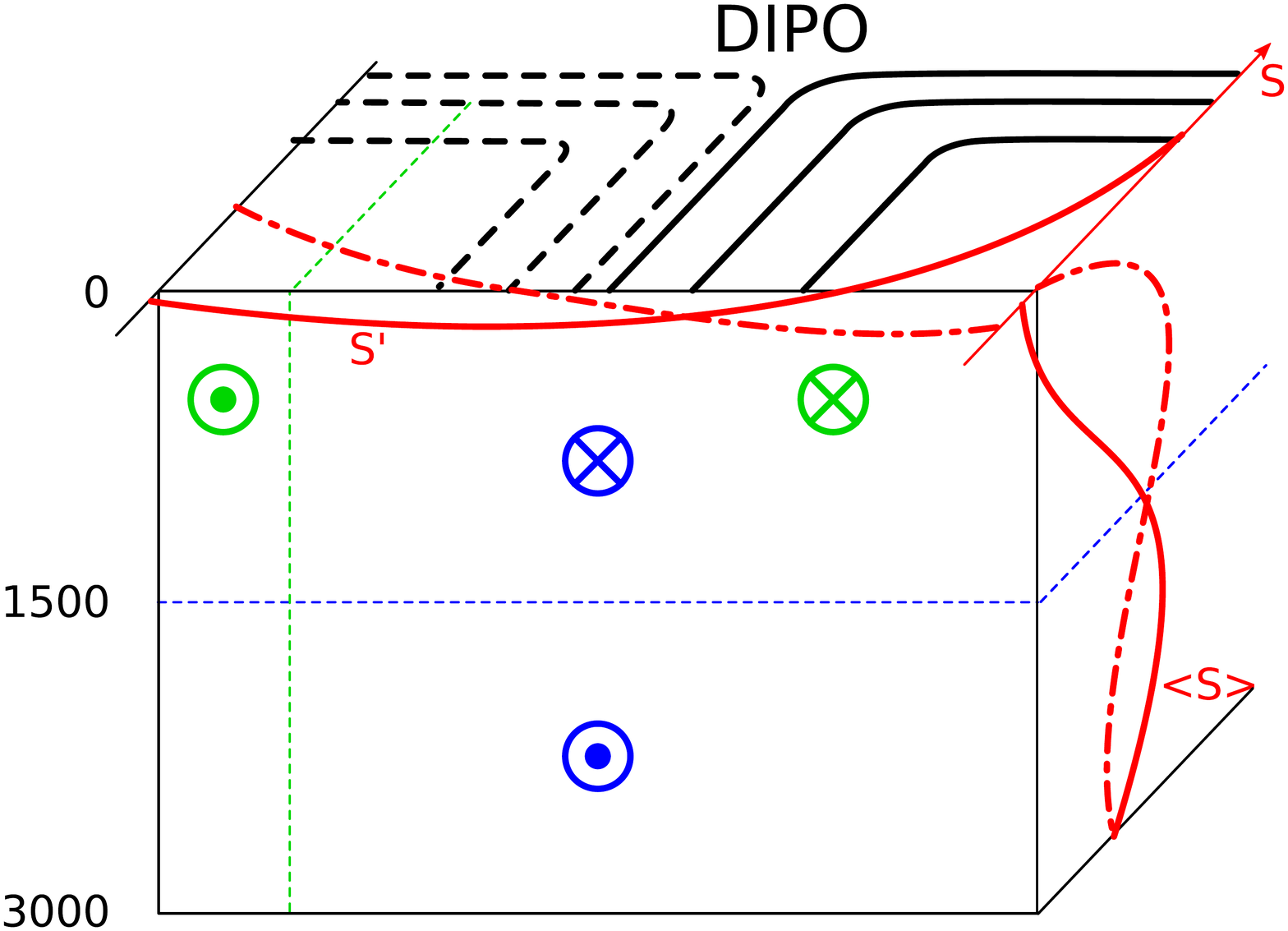}
    \end{center}
    \caption{
    A sketch showing the effect of the dipole anomaly on the salinity at the southern border of the Atlantic Ocean.
    A vertical transect is shown.
    The blue colour refers to the overturning circulation, with inflow of intermediate waters above $\approx1,500\,\mathrm{m}$ and outflow below.
    The green colour refers to the azonal circulation, with outflow in the western boundary and inflow elsewhere.
    The dipole anomaly DIPO (with strength $\delta_p$) is shown as contours with full lines representing freshwater input. 
    The zonal ($\left<S\right >$) and azonal ($S'$, vertically averaged) profiles of salinity are shown as red lines.
    The full and dashed red lines refer to the ``biased'' ($\delta_p = 0$) and ``unbiased'' ($\delta_p > 0$) situations, 
    respectively.  $\delta_p$ reverses the zonal salinity contrast (increasing $M_{az}$) but also increases the salinity 
    of the intermediate waters, leading to a decrease in  $M_{ov}$.
    \label{fig:sketch}
    }
  \end{figure*}

  Based on the results of these experiments, we can identify different regions in the parameter space, with markedly different sensitivity of AMOC to external perturbations.
  These regions could serve as a guide for assessing the AMOC sensitivity in different climate models.
  Starting point is the estimation of the equivalent freshwater budget for the Atlantic Ocean and in particular the coupling between the $M_{ov}$ and $M_{az}$ in the freshwater budget.
  As sketched in  figure~\ref{fig:sketch}, a positive value of $\delta_p$ induces a fresher eastern part of the South Atlantic which causes an increase in the azonal freshwater transport $M_{az}$.
  Keeping other terms constant, an increase in $M_{az}$ is compensated by a decrease in $M_{ov}$ leading to more freshwater export out of the Atlantic by the AMOC due to changes in the zonally averaged salinity profile.

  The freshwater budget gives direct hints to the stability properties of the overturning circulation in the model \cite{huisman_indicator_2010} and provides an estimate of the position of the model in the regime diagram of Fig.~\ref{fig:THCM}.
  A dipole anomaly can move the system closer to the regime of multiple equilibria, as the second bifurcation point $L_2$ (green line in Fig.~\ref{fig:THCM}) moves towards smaller values of $\gamma_p$ for increasing values of the dipole anomaly.
  This is, incidentally, a clear indication of the fact that a model with a correct zonal salinity contrast is likely to be closer to the multiple equilibria region as well.
  From experiment Crev it can be inferred that a $\delta_p=0.5\Sv$ anomaly is able to extend the multiple equilibria area almost to the $\gamma_p=0\Sv$  line.

  Even if the values of $\delta_p$ and $\gamma_p$ needed to collapse the model, or even reach the multiple equilibria regime, are obviously model--dependent, the anomalies associated with these transitions in SPEEDO can provide an efficient guide to demonstrate whether ME of the AMOC exist in other climate models.

  Our experiments are based on artificially imposing changes to surface freshwater fluxes but we argue that similar results may hold when the surface fluxes change due to either physical processes as global warming or modifications in a climate model aimed at e.g. correcting biases in surface fluxes.
     Regarding the latter point, underestimation of stratocumulus cover at low latitudes is a well known issue in numerical models~\cite{karlsson_cloud_2008} that is very likely to have a strong impact on surface salinity in the south eastern Atlantic.
  Such a model bias, and its correction, would lead to anomaly patterns very similar to the dipole used in the present work.
  Also Agulhas leakage, its misrepresentation in ocean models, in particular at low resolution, and possible changes in it due to natural or human--induced variability~\cite{beal_role_2011} are likely to produce changes in the freshwater budget of the Atlantic Ocean.
  Even if only at an idealised level, our results may be representative of such changes.
  Our demonstration of the strong dependence of the AMOC sensitivity to the details of the freshwater budget may also contribute to solve the contradiction between the modest ice losses expected from Greenland ice sheet ($\mathcal{O}(0.01\Sv)$ according to~\textcite{van_den_broeke_partitioning_2009}, in connection with present day warming) and the common attribution of abrupt paleoclimate variability to AMOC collapse and recovery, which needs substantially stronger perturbations in climate models (see for instance~\textcite{timmermann_coherent_2003} and references therein).

  The findings presented here have great relevance for present state of the art climate models.
  The biases in salt transport by the overturning circulation found by~\textcite{drijfhout_stability_2011} are probably connected with biases in the salinity field of the models.
  Our arguments, involving large spatial and temporal scales and the simple physical mechanism of salt--advection feedback, are unlikely to be affected by changes in resolution, or by different parametrisations.
  The results of~\textcite{hawkins_bistability_2011} and~\textcite{de_vries_atlantic_2005} hint at the possibility that $M_{ov}$ could be a reliable indicator of an approaching collapse of the AMOC.
  It is anyway of fundamental importance to carefully consider all the different components of the freshwater budget, as different models may have a markedly different behaviour for some of the terms that were not considered in this work (e.g.~the magnitude of Bering Strait transport may be non negligible, the deep overturning between the Southern Ocean and the Atlantic may give a substantial contribution to $M_{ov}$ and relative importance of advective and diffusive terms may change the compensating behaviour of $M_{ov}$ and $M_{az}$).
  It is also important to stress that there is no theoretical background, at present, supporting the use of $M_{ov}$ as a stability indicator far from the steady state, so that great care should be taken when analysing transient experiments as done by~\textcite{hawkins_bistability_2011}.
  We also can not exclude that simulations at higher resolution, or with more sophisticated parametrisation, may uncover the importance of other processes that could not be taken into account in the present work.
  In the work of~\textcite{farneti_role_2010}, as an example, the mesoscale eddies provide a sink to the potential energy produced by the Ekman flux in the Southern Ocean that would otherwise be absent at lower resolution, markedly changing the response of the ocean to wind stress variability.
  Whether this, or similar processes, are relevant for the stability of the AMOC is at the moment unclear.
  
  A correct representation of salinity, and consequently of the equivalent freshwater fluxes, is needed if any inference or prediction on the stability of the AMOC has to be drawn from model results.
  Changes in the surface fluxes, either physically motivated or artificially imposed as in the case of this work, can change entirely the response of the overturning circulation to an identical perturbation.
  A key result of this study is that the stability of the AMOC crucially depends on salinity/freshwater anomalies in the South Atlantic.
  This result challenges the traditional view of the stability of the AMOC being solely determined by processes in the North Atlantic.

  \begin{acknowledgments}
  A.A.C. acknowledges the Netherlands Organization for Scientific Research (NWO) for funding in the ALW program.
  M.d.T. acknowledges NWO for funding through the TopTalent Grant.
  \end{acknowledgments}

  \appendix
  \section{Equivalent freshwater budget in the Atlantic Ocean}
  \label{sec:Appendix}

   The volume budget of the Atlantic Ocean can be written as:
   \begin{equation}
      \frac{\partial V}{\partial t} = \int_{BS}v\,\dd x \dd z + \int_{30S} v \,\dd x \dd z - EPR, \label{eq:Volume_eq}
   \end{equation}
   where $v$ is the meridional velocity, $BS$ and $30S$ indicate integration over a zonal transect in the Bering strait and at $30\Sl$ in the Atlantic Ocean respectively.
   $EPR$ is the net evaporation over the basin and $V$ is volume.
   The subscript $t$ denotes time derivative.
   The volume change in the basin is the balance between inflow from zonal boundaries and net evaporation.

   \subsection{Equivalent freshwater budget}

   Local salt conservation is expressed by:
   \begin{equation}
      \frac{\partial S}{\partial t} + \mathbf{u}\cdot{\nabla} S = -\nabla \mathcal{F}_S, \label{eq:salt_evolution}
   \end{equation}
   where $S$ is salinity, $\mathbf{u}$ is the horizontal velocity vector and $\mathcal{F}_S$ is the diffusive salt flux.
   Diffusive fluxes depend on model definition, so we do not write explicitly here.
   Integrating Eq.~\ref{eq:salt_evolution} over the whole Atlantic basin, using Gauss theorem and assuming no salt flux at the surface and bottom and no diffusion across Bering strait, one obtains:
   \begin{equation}
      \frac{\partial}{\partial t}\int_{Atl}S \,\dd V - \int_{30S}vS \,\dd x \dd z - \int_{BS}vS \,\dd x \dd z = \int_{30S} \mathcal{F}_S \,\dd x \dd z. \label{eq:budget}
   \end{equation}
   Using a reference salinity $S_0$:
   \begin{equation}
    S_0 = \frac{\int_{30S}S\,\dd x \dd z}{\int_{30S}\,\dd x \dd z}, \label{eq:S_0}
   \end{equation}
   an equivalent freshwater transport can be defined from~\ref{eq:budget}:
   \begin{equation*}
      -\frac{1}{S_0}\frac{\partial}{\partial t}\int_{Atl}S \,\dd V  = -\frac{1}{S_0} \int_{30S}vS \,\dd x \dd z -\frac{1}{S_0} \int_{BS}vS \, \dd x \dd z  - \frac{1}{S_0} \int_{30S} \mathcal{F}_S \, \dd x \dd z,
   \end{equation*}
   written in short as:
   \begin{equation}
      Q_t = M_{30S} -\frac{1}{S_0} \int_{BS}vS \, \dd x \dd z + M_{d} + Res,  \label{eq:freshwater_budget}
   \end{equation}
   where $Res$ is a residual.

   \subsection{Virtual freshwater transport at $30^\circ S$}

   The term $M_{30S}$ in Eq.~\ref{eq:freshwater_budget} is split in two parts:
   \begin{equation}
      \begin{split}
          M_{30S} = -\frac{1}{S_0} \int_{30S}vS \,\dd x \dd z & = -\frac{1}{S_0} \int_{30S}(\langle v \rangle + v')(\langle S \rangle + S') \,\dd x \dd z \\
          &= -\frac{1}{S_0} (\int_{30S}\langle v \rangle \langle S \rangle \,\dd x \dd z + \int_{30S}v'S' \,\dd x \dd z), \label{splitting}
      \end{split}
   \end{equation}
  where, for a generic field $f$, the zonal operator is $\left<f\right>=\int\limits_{60\Wl}\limits^{20\El} f \dd x/\int \dd x$ and the azonal operator is $f'=f-\left< f \right >$.

   Using the two definitions:
   \begin{equation*}
      \begin{split}
          M_{ov} &= -\frac{1}{S_0} \int_{30S}\tilde{v} (\langle S \rangle - S_0) \,\dd x \dd z \\
          M_{az} &= -\frac{1}{S_0} \int_{30S}v'S' \, \dd x \dd z,
       \end{split}
   \end{equation*}
  where for a generic field $f$ the barotropic operator is $\overline{f} = \int f \dd z / \int \dd z$ and the baroclinic operator is $\tilde{f}=f-\overline{f}$, Eq.~\ref{splitting} becomes:
   \begin{equation}
    M_{30S} = M_{ov} + M_{az} - \int_{30S}v\,\dd x \dd z. \label{eq:T_30S_short}
   \end{equation}

   Using Eq.~\ref{eq:T_30S_short}, the volume \textit{outflow} of water at $30S$ can be represented by a virtual freshwater \textit{inflow}.

   Putting together Eqs.~\ref{eq:Volume_eq}~and~\ref{eq:T_30S_short} into the virtual freshwater budget of Eq.~\ref{eq:freshwater_budget} we obtain Eq.~\ref{eq:EPR}:
   \begin{equation*}
    EPR + Q_t + V_t = M_{ov} + M_{az} + M_d + M_{BS} +Res,
   \end{equation*}
   with $V_t=\partial V/\partial t$ the volume drift in the basin, and defining $M_{BS}= \int_{BS}v\,\dd x \dd z -\frac{1}{S_0} \int_{BS}vS \, \dd x \dd z$.
   The inflow of freshwater in the basin must balance net evaporation and the drift in volume and salinity.

   In the calculations, baroclinic velocity can be used instead of actual one, as long as Eq.~\ref{eq:S_0} is used as the definition of $S_0$.
   This stems from the definition of baroclinic velocity.

   \subsection{Freshwater budget in CLIO}

   The terms in Eq.~\ref{eq:EPR} for the control state of the EMIC SPEEDO are reported in table~\ref{tab:freshwater}.
   Very similar numbers are obtained for the HCM.
   In the model, the freshwater anomalies are implemented as a virtual salt flux, which can easily be accounted for by a surface salt flux in Eq.~\ref{eq:salt_evolution}.
   Here we include this term into $EPR$ for simplicity.

   The largest terms in the budget are $EPR$, $M_{ov}$ and $M_{az}$ by at least one order of magnitude.
   Moreover, the other terms are also less sensitive to the freshwater anomalies applied so that they are approximately constant, with the exception of $Q_t$.
   As discussed in section~\ref{sec:discussion} this depends on the details of the model used.
   As an example, in THCM, the term $M_d$ plays instead a primary role, and behaves in a way similar to $M_{az}$ in SPEEDO.


  \clearpage

  \singlespacing

  \begin{table}[htbp]
    \centering
    \begin{tabular}{p{1.5cm} | p{1.3cm} p{2.0cm} p{2.0cm}}
      Name & Model & $\gamma_p\;[\Sv]$ & $\delta_p\;[\Sv]$\\
      \hline
      A & HCM & $0.0-0.4$ & $0.0-0.0$ \\
      B & HCM & $0.0-0.0$ & $0.0-0.5$ \\
      C & HCM & $0.0-0.4$ & $0.5-0.5$ \\
      Crev & HCM & $0.4-0.0$ & $0.5-0.5$ \\
      D & HCM & $0.4-0.4$ & $0.0-0.5$ \\
      Drev & HCM & $0.4-0.4$ & $0.5-0.0$ \\
      HCMc & HCM & $0.0-0.4$ & $0.0-0.5$ \\
      GCMc  & EMIC & $0.0-0.4$ & $0.0-0.5$ 
    \end{tabular}
    \caption{List of the experiments with the abbreviations used in the text.
    The initial and final values of the two freshwater anomaly amplitude are listed.
    The anomalies change linearly in time during the experiment.
    All experiments last 12,000 years with the exception of HCMc and GCMc lasting 4,000 years.}
    \label{tab:exps}
  \end{table}

  \begin{table}[htbp]
    \centering
    \begin{tabular}{p{1.5cm} | p{3.5cm}}
       $EPR$    & $6.1\cdot 10^{-1}\Sv$ \\
       $M_{ov}$ & $3.2\cdot 10^{-1}\Sv$ \\
       $M_{az}$ & $1.5\cdot 10^{-1}\Sv$ \\
       $M_d$    & $3.0\cdot 10^{-2}\Sv$ \\
       $M_{BS}$ & $6.1\cdot 10^{-2}\Sv$ \\
       $Q_t$    & $-4.0\cdot 10^{-5}\Sv$ \\
       $V_t$    & $-1.3\cdot 10^{-3}\Sv$ \\
       $Res$    & $4.7\cdot 10^{-2}\Sv$
    \end{tabular}
    \caption{Values of the different terms in Eq.~\ref{eq:EPR} as computed in the control run of the EMIC SPEEDO. The budget is computed from yearly data, using a 1,000 years long simulation at statistical steady state.
    \label{tab:freshwater}}
  \end{table}

\end{document}